# Enforced Interface Constraints for Domain Decomposition Method of Discrete Physics-Informed Neural Networks

Jichao Yin[ad]; Mingxuan Li[c]; Jianguang Fang[b*]; Hu Wang[ad*]

[a] *State Key Laboratory of Advanced Design and Manufacturing Technology for Vehicle, College of Mechanical and Vehicle Engineering, Hunan University, Changsha, 410082, P.R.China*

[b] *University of Sydney, School of Aerospace, Mechanical and Mechatronic Engineering, Sydney, Australia*

[c] *State Key Laboratory of Precision Electronics Manufacturing Technology and Equipment, Guangdong University of Technology, Guangzhou, 510006, P.R.China*

[d] *Beijing Institute of Technology Shenzhen Automotive Research Institute, Shenzhen, 518000, P.R.China*

**Keywords**: This study introduces a discrete physics-informed neural network (dPINN) framework augmented with enforced interface constraints (EIC) for modeling physical systems under the domain decomposition method (DDM). Built upon finite element-style mesh discretization, the dPINN accurately evaluates system energy through Gaussian quadrature-based element-wise integration. To ensure physical field continuity across subdomain interfaces, the EIC mechanism enforces interfacial displacement constraints without requiring auxiliary sampling or loss penalties. This formulation supports independent meshing in each subdomain, simplifying preprocessing and improving computational flexibility. Additionally, by eliminating the influence of weak spatial constraints (WSC) commonly observed in traditional PINNs, the EIC-dPINN delivers more stable and physically consistent predictions. Extensive two- and three-dimensional numerical experiments validate the proposed framework's accuracy and demonstrate the computational efficiency gains achieved through parallel training. The results highlight the framework's scalability, robustness, and potential for solving large-scale, geometrically complex problems.

---

\* Corresponding author. E-mail address: wanghu@hnu.edu.cn (H. Wang)

# 1. Introduction

In the rapidly evolving field of the scientific and engineering computation, Physics-Informed Neural Networks (PINNs) [1,2] have emerged as a powerful computational approach, demonstrating unique advantages and significant potential. By integrating fundamental physical laws into the learning process of neural networks, PINNs not only enable accurate predictions in scenarios with limited data but also effectively address complex problems [3,4]. This integration offers a novel perspective and a valuable tool for advancing scientific research. However, as computational challenges grow in scale and complexity, leveraging Physics-Informed Neural Networks (PINNs) to tackle large-scale, real-world problems has become a pressing research priority. To address this challenge, Domain Decomposition Methods (DDMs) [5–7] provide a promising approach by partitioning complex computational domains into smaller, more tractable subdomains. This decomposition enhances parallel computing efficiency, accelerates computations, and facilitates the simulation of large-scale systems. When coupled with PINNs [8,9], DDMs significantly expand the scope of PINNs, enabling them to model a wider range of intricate physical phenomena.

DDM was initially incorporated into the PINNs framework by jagtap et al. [8] to alleviate computational burdens and enhance solution accuracy for complex physical systems. A notable advancement in this direction is the development of the conservative PINN (cPINN), which enforces conservation laws across subdomain interfaces. By partitioning the solution domain and introducing flux-based regularization terms at interface collocation points, cPINN ensures inter-subdomain continuity and enables parallel computations, thus leveraging multi-CPU and multi-GPU architectures for improved scalability. Subsequently, the Extended PINN (XPINN) [9] was later introduced to extend DDMs to irregular and non-convex geometries. The effectiveness of both cPINN and XPINN in distributed computing environments was empirically validated by Shukla et al. [10]. However, these methods fundamentally rely on loss minimization to enforce interfacial continuity, which may not guarantee strict adherence to physical constraints, potentially leading to solution inconsistencies [11]. To address temporal complexity, Meng et al. [12] proposed the Parareal PINN (PPINN), inspired by the Parareal time-parallel integration framework. PPINN decomposes the temporal domain into shorter intervals and employs a hierarchical architecture comprising a coarse-grained PINN for initialization and a fine-grained PINN for iterative refinement. This strategy significantly accelerates training while maintaining temporal accuracy. Further efforts to improve solution consistency include the hp-Variational PINN (hp-VPINN) introduced by Kharazmi et al. [13], which draws on Finite Element Method (FEM) principles. By combining spatial (h-) decomposition with localized learning via parameter (p-) decomposition, and employing a global potential function, hp-VPINN enforces solution continuity across subdomains. Li et al. [14] advanced this framework by proposing a Fourier Feature-based deep domain decomposition method that integrates spectral embeddings and overlapping subdomain boundaries to better capture high-frequency solution features.

Additionally, Extreme Learning Machines (ELMs) [15] have gained traction in DDM frameworks due to their rapid training and low computational overhead. Researchers such as Dwivedi [16] and Dong [17] utilized ELMs to construct lightweight subnetworks within subdomains, achieving notable efficiency improvements. However, the fixed hidden-layer parameters in ELMs restrict model expressivity, limiting their adaptability in solving high-dimensional nonlinear problems. Although comparable to FEM in efficiency, their applicability to more complex scenarios remains limited. To address these limitations, Hu et al. [18] proposed a novel DDM approach employing a gating network to perform weighted summation over outputs of multiple subnetworks defined across the entire domain. Unlike conventional DDMs restricted to local subdomains, this full-domain architecture enables dynamic reallocation of model capacity across regions, enhancing both stability and accuracy of the composite solution.

The integration of domain decomposition techniques into PINNs has notably advanced computational efficiency and parallelization, thereby facilitating the solution of large-scale, complex problems. Nonetheless, the challenge of discontinuities at subdomain interfaces persists as a significant obstacle, limiting the broader applicability of DDM based-PINNs. Similar to the DDMs, the traditional non-conforming FEM [19] performs field simulations on meshes tailored to different subdomains. This approach has been widely applied in scenarios such as multi-physics coupling, complex structural analysis, and heterogeneous domain interfacing [20,21]. A key characteristic of this method is that the boundary meshes between adjacent subdomains are typically non-matching, which can result in node misalignment, geometric gaps, or local overlaps at the interface. These inconsistencies violate the fundamental assumptions of geometric and solution continuity in classical FEMs [22–24]. To address such non-conformity, conventional assembly strategies based on nodal consistency and interpolation continuity are no longer directly applicable. Instead, weak continuity conditions or interface coupling techniques must be employed to ensure accurate transfer of physical quantities across subdomains. Representative methods include the multi-point constraints (MPC) method [25] and the Nitsche method [26], as well as several others. These approaches provide effective means of numerical coupling between non-conforming meshes, each offering different trade-offs among accuracy, stability, and computational cost.

Inspired by the concept of MPC in nonconforming FEMs [27,28], this work develops a discrete PINN (dPINN) framework on a mesh-based structure. The key idea is to perform energy evaluation at the level of discrete mesh elements, treating each element as the minimal unit for energy integration, which is then assembled to yield the total energy of the system. In contrast to classical MPC formulations, where continuity across nonconforming interfaces is enforced via explicit constraints on the mapped displacements of adjacent subdomain nodes, the proposed method circumvents the need for such enforcement. Instead, it achieves inter-subdomain continuity by interpolating the displacements of nodes on the interface of neighboring subdomains using the nodal values from the adjacent elements. An additional advantage of this approach is that it naturally accommodates non-matching meshes at subdomain interfaces, thereby relaxing the meshing requirements for each subdomain. This flexibility makes the method particularly suitable for handling geometrically complex problems such as stress concentration [29]. This interpolation-based mechanism ensures displacement continuity across subdomain boundaries without the need to introduce additional flux computations or penalty terms [8,9]. By avoiding the inclusion of explicit interface losses or auxiliary physical quantities, the proposed approach significantly reduces the complexity of domain decomposition. Furthermore, it mitigates the adverse effects of loss imbalance often observed in traditional PINNs [30,31], thereby enhancing both the convergence behavior and training efficiency of the model.

The following was a description of the organization of the paper. In Section 2, xx. In Section 3, xx. Finally, xx in Section 4.

## 2. The methodologies

### 2.1 Discrete physics-informed neural networks

For standard PINNs, the primary objective is to solve partial differential equations (PDEs), which can generally be written in the following form [8]:

$$\begin{cases} \mathcal{F}(\boldsymbol{u}(\boldsymbol{x}); \boldsymbol{\lambda}) = f(\boldsymbol{x}), & \boldsymbol{x} \in \Omega \subset \mathbb{R}^d \\ \mathcal{B}_k(\boldsymbol{u}(\boldsymbol{x})) = g_k(\boldsymbol{x}), & \boldsymbol{x} \in \Gamma_k \subset \partial\Omega \end{cases}, \quad (1)$$

where $\mathcal{F}(\cdot)$ denotes the differential operator, $\boldsymbol{u}(\boldsymbol{x})$ represents the target field variable to be inferred, $\boldsymbol{\lambda}$ corresponds to the parameters of the PDE, $f(\boldsymbol{x})$ is a known source term, $\boldsymbol{x}$ denotes spatial points within the solution domain $\Omega$. To ensure a well-posed problem, the boundary conditions $\mathcal{B}_k(\cdot)$ are imposed on the domain boundary $\Gamma_k$, with the prescribed boundary values $g_k(x)$. In the context of solid mechanics, the governing PDE system corresponds to the equilibrium equation under static conditions [32], formulated as:

$$\begin{cases} \nabla \sigma + b = 0, & \text{in } \Omega \\ u = \bar{u}, & \text{in } \Gamma_D, \\ \sigma n = \bar{t}, & \text{in } \Gamma_N \end{cases} \quad (2)$$

where $\sigma$ is the Cauchy stress as a function of the displacement field $u$, and $b$ is the body force. The boundary conditions are specified by displacement (Dirichlet) conditions $\bar{u}$ applied on the boundary $\Gamma_D$ and traction (Neumann) conditions $\bar{t}$ imposed on the boundary $\Gamma_N$. These two types of boundary conditions are mutually exclusive and do not overlap. To facilitate numerical approximation, the strong form (2) can be reformulated into a variational (weak) form using the principle of virtual work, the weak form is expressed as:

$$\int_\Omega (\delta \varepsilon^T \sigma - \delta u^T b)dx - \int_{\Gamma_N} \delta u^T \bar{t} ds = 0. \quad (3)$$

where $\delta \varepsilon$ is the virtual strain obtained from the virtual displacement $\delta u$ (test function), and the strain-displacement relationship is given by:

$$\varepsilon = \frac{1}{2}(\nabla u^T + \nabla u). \quad (4)$$

Most PINN frameworks formulate their loss functions based on one of two fundamental principles. The first approach directly enforces the governing PDEs, as presented in Eq. (2), leading to a strong-form residual formulation. The second approach derives from the variational principle, as expressed in Eq. (3), which yields an energy-based loss function grounded in the principle of minimum potential energy [33].

$$\mathcal{L}_{PDE} = \frac{1}{Nf}\sum_{i=1}^{Nf} \left| \mathcal{F}(u_\theta(x_i); \lambda) - f(x_i) \right|^2 + \mathcal{L}_{DDM};$$
$$\mathcal{L}_{ENG} = \Pi(u_\theta(x_i)) \approx \frac{1}{Ne}\sum_{i=1}^{Ne} (\sigma_\theta(x_i)\varepsilon_\theta(x_i)^T - \bar{t}(x_i)u_\theta(x_i)^T) + \mathcal{L}_{DDM}, \quad (5)$$

where $u_\theta$ is the displacement predicted by PINNs, and $\mathcal{L}_{DDM}$ is an additional loss term for DDM, typically consisting of multiple components. Whether the loss function is based on the PDE or on energy principles, it is computed at the collocation points $x_i$, with the corresponding loss terms calculated to accurately approximate the global continuous solution. However, loss functions constructed using collocation points place high demands on the sampling strategy, as both the number and distribution of samples can significantly affect the prediction accuracy of the PINN. Furthermore, the additional loss terms $\mathcal{L}_{DDM}$ are introduced to enforce continuity across subdomains, ensuring solution consistency but also increasing the computational complexity and training difficulty of the PINN.

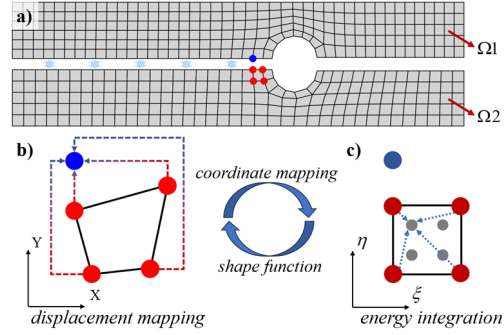

Figure 1 Schematic diagram of multi-subdomain energy integration: (a) node–element pairing at the interface, (b) displacement mapping in the physical space, and (c) Gaussian integration in the reference space.

To address this challenge, an energy evaluation framework for PINN inspired by the FEM is proposed, grounded in mesh-based discretization (as presented in Figure 1 (a)). In this formulation, the energy is computed over a collection of non-overlapping discrete elements that together fully cover the computational domain. These energy evaluations are subsequently employed to construct the loss function. This approach ensures that the energy assessments at each training iteration remain both physically consistent and numerically accurate.

$$\mathcal{L} = \frac{1}{2}\sum_{e=1}^{Ne}\sum_{g=1}^{Ng}\omega_g \boldsymbol{\sigma}_\theta(\boldsymbol{x}_g)\boldsymbol{\varepsilon}_\theta(\boldsymbol{x}_g)^{\mathrm{T}} - \sum_{d=1}^{Nd}\overline{\boldsymbol{t}}(\boldsymbol{x}_d)\cdot\boldsymbol{u}_\theta(\boldsymbol{x}_d). \tag{6}$$

The Eq. (6) demonstrates the formulation of the loss function based on strain energy and external work, corresponding to the energy method. The first term computes the strain energy, where the strain energy of each element is evaluated through a weighted summation at the Gauss integration points, $\boldsymbol{x}_g$ [34]. The total strain energy of the system is obtained by accumulating the contributions from all $Ne$ discrete elements. The second term represents the external work, calculated as the summation of the product of the applied loads at the boundary and the corresponding displacements at the $Nd$ loading locations, $\boldsymbol{x}_d$.

Notably, the loss function provided in Eq. (6) ultimately simplifies to a single term, namely the potential energy. In contrast to the two loss function forms presented in Eq. (5), there is no necessity to introduce an additional interface loss term. Analogous to prescribed displacement constraints [35], tthe interface penalty terms are reformulated as enforced displacement conditions. This transformation constitutes a significant innovation in the present study, which will be explored in detail in the following sections.

## 2.2 Domain decomposition method with enforced interface constraints

To improve computational efficiency and alleviate the burden on computational hardware, the DDM technique, widely used within FEM, has been subsequently extended to PINNs. However, the application of DDM within PINN necessitates the introduction of additional terms in the loss function, which not only fails to rigorously ensure boundary continuity but also increases the complexity of training the PINN. To address this issue, an enforced interface constraints-based DDM is proposed, which ensures strict boundary continuity without the need for additional loss terms.

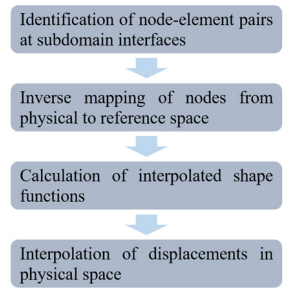

Figure 2 Interface constraint preprocessing flow

As depicted in Figure 1 (a), the solution domain is subdivided into two subdomains. To ensure the continuity of the displacement field across the subdomains, a mapping relationship is established between the elements at the subdomain boundaries and the adjacent interface nodes of the neighboring subdomain. The mapping can be either unidirectional or bidirectional. For the sake of simplicity, a unidirectional mapping is used as the illustrative example.

A node located on the interface of subdomain $\Omega 1$ is first designated as a reference point (denoted by a blue dot). Using a distance-based criterion within the physical space, the nearest element from an adjacent subdomain $\Omega 2$ (represented by the red dot) is identified. This procedure establishes a comprehensive set of node–element pairs at the interface, which forms a critical data structure for subsequent interface coupling analyses, facilitating efficient data transfer between subdomains.

To ensure consistency between the strain energy integration domain and the displacement interpolation domain at the interface, these node–element pairs must be fully mapped into the reference space. This mapping guarantees compatibility of the shape function interpolation. In the reference space, as shown in Figure 1 (a) and (c), elements are represented as regular squares. However, the coordinates of the nodes in this space cannot be directly inferred from their physical counterparts and must be determined by solving the following equation:

$$R(\xi) = \sum_{i=1}^{m} N_i(\xi,\eta)x_i - x_o = 0, \tag{7}$$

where node $\xi = (\xi,\eta)$ in the reference space is the mapped coordinates of node $x_o$ at physics space. $N_i$ are the shape functions, $x_i$ are the coordinates associated with the vertices of the element in physical space, and $m$ is the number of the vertices. This equation is typically solved using the Newton–Raphson method [36],

$$\xi^{(k+1)} = \xi^{(k)} - \left(\frac{\partial R(\xi^{(k)})}{\partial \xi}\right)^{-1} R(\xi^{(k)}), \tag{8}$$

with the Jacobian matrix given by,

$$\frac{\partial R(\xi^{(k)})}{\partial \xi} = \sum_{i=1}^{m} \frac{\partial N_i(\xi,\eta)}{\partial \xi} x_i = J. \tag{9}$$

The initial guess for $\xi^{(0)}$ is usually set at the center of the reference element, and iterations proceed until convergence based on a predefined criterion, i.e., $\|R(\xi^{(k)})\| < \tau$.

Once the inverse-mapped coordinates are obtained, the corresponding shape functions at each element vertex can be evaluated as:

$$N_i^e(\xi^{(k)},\eta^{(k)}) = \frac{1}{4}(1+\xi_i\xi^{(k)})(1+\eta_i\eta^{(k)}), \tag{10}$$

where $e$ represents the index of node-element pair, and $\xi_i = (\xi_i,\eta_i)$ refers to the coordinates of the element's vertices in the reference space.

These shape functions are then used in the physical space to interpolate the displacements at the interface nodes. The interpolated displacement values are computed by combining the nodal displacements of the element's vertices with the evaluated shape functions, enabling accurate displacement estimation across subdomain interfaces.

$$u^e = \sum_{i=1}^{m} N_i^e(\xi^{(k)},\eta^{(k)}) x_i. \tag{11}$$

It is important to note that all preprocessing steps preceding the displacement mapping (Eq. (11)) are conducted before the initiation of PINN training and do not require repetition during the training process. Consequently, the computational overhead associated with these preprocessing steps remains negligible in comparison to the total training cost. In contrast, displacement mapping must be executed at each training epoch, with the mapped displacements replacing the corresponding predictions generated by the neural network.

## 2.3 Parallel Computing Scheme

Building upon the proposed methodologies, a parallel computational framework for PINNs is developed, integrating both data parallelism and model parallelism to enable concurrent execution across multiple GPU devices. This design facilitates significant acceleration in model training while maintaining flexibility and scalability. The overall workflow of the parallel computing strategy is illustrated in Figure 3. The solution domain is first discretized and partitioned into a set of non-overlapping subdomains. Notably, the mesh partitioning process imposes no strict requirements on the conformity of nodes at subdomain interfaces. The framework accommodates non-conforming meshes at interfaces, allowing for potential overlaps and discontinuities.. This relaxed constraint greatly simplifies the mesh partitioning process by allowing each subdomain to be meshed independently, thereby reducing meshing complexity and computational overhead. For convenience of notation, the mesh associated with each subdomain is represented symbolically $\mathcal{M}^{(i)}$.

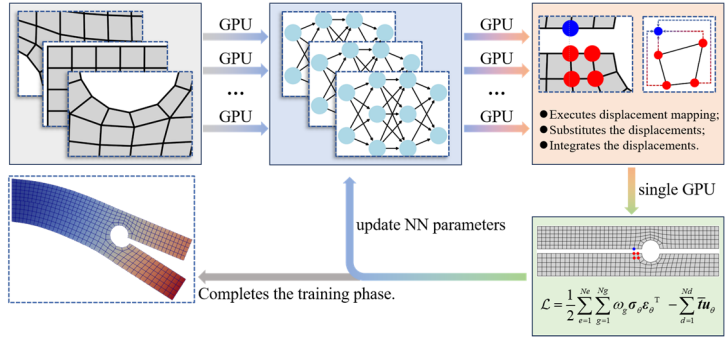

Figure 3 Illustration of the multi-GPU parallel computing scheme with enhanced interface constraint conditions.

To realize multi-GPU parallelism, multiple neural networks $\mathcal{NN}^{(i)}$ are deployed independently on distinct GPU devices. The employed network architecture in this study is consistent across models, with the spatial coordinates serving as inputs. These inputs are first transformed using random Fourier features $f^{rff}$ [37], and subsequently propagated through multiple linear hidden layers $f^L$ of uniform width. Each hidden layer is followed by a normalization layer $f^N$ [38] and a hyperbolic tangent (Tanh) activation function to introduce nonlinearity. The network ultimately outputs the predicted displacement field.

$$\begin{cases} \mathcal{NN}^{(i)}(\boldsymbol{x},\boldsymbol{\theta}^{(i)}) = f^{out} \circ f^H \circ ... \circ f^H \circ f^{rff}(\boldsymbol{x}) \\ f^H = \text{Tanh}(f^N \circ f^L) \end{cases}. \quad (12)$$

The networks can further be customized to the characteristics of their respective subdomains through adjustments to architectural parameters such as depth, width, and activation functions. The number of neural networks corresponds directly to the number of subdomains, and each mesh–network pair is co-located on the same GPU to minimize data transfer overhead. Furthermore, the mesh data $\mathcal{M}^{(i)}$ for each subdomain remains fixed throughout training, and the association between each $\mathcal{NN}^{(i)}$ and its corresponding $\mathcal{M}^{(i)}$ is also constant. This fixed mapping enhances the stability and consistency of the training process. Since each $\mathcal{NN}^{(i)}$ operates only on a local subset of the global mesh, the volume of input data per model is significantly reduced. As a result, both forward computation time and memory consumption are greatly optimized. This not only accelerates the training process but also enables the framework to efficiently handle larger-scale problems with finer discretization or more complex geometries.

While the strain energy within each subdomain can be evaluated independently, the formulation of the loss function necessitates access to the global strain energy of the entire system. Consequently, displacement predictions from all neural networks must be gathered and transferred to a centralized device for unified processing.

$$\begin{cases} \boldsymbol{u}_{lc}^{(i)} = f(\mathcal{NN}^{(i)}(\mathcal{M}^{(i)})) \\ \boldsymbol{u}_{\theta} = \Sigma \boldsymbol{u}_{lc}^{(i)} \end{cases}, \tag{13}$$

where $f(\cdot)$ denotes the displacement mapping operator (as previously defined in Eq. (11)), which overlays the mapped displacements onto the corresponding predictions generated by the $\mathcal{NN}^{(i)}$. The symbol $\boldsymbol{u}_{lc}^{(i)}$ represents the local displacement solution within each subdomain, while $\Sigma$ refers to the process of migrating the displacement data from multiple GPUs to a centralized device and assembling them into the complete displacement field $\boldsymbol{u}_{\theta}$ over the global solution domain. This reconstruction procedure ensures that the locally predicted displacements are coherently integrated into a unified representation spanning the entire computational domain. Following the application of these operators, both the global strain energy and the external work are evaluated on a single GPU. This centralized computation facilitates the construction of the physics-informed loss function, as previously defined in Eq. (6).

From an implementation perspective, displacement hard constraint [39] is typically enforced after the full displacement field has been assembled (see Eq. (13)), followed by the evaluation of the system's energy functional (see Eq. (6)).

$$\boldsymbol{u} = g(\boldsymbol{x}) + l(\boldsymbol{x})\boldsymbol{u}_{\theta}(\boldsymbol{x}), \tag{14}$$

where the function $g(\boldsymbol{x})$ is introduced to inherently satisfy the Dirichlet boundary condition, while the function $l(\boldsymbol{x})$ serves to enforce zero values at the nodes located on the Dirichlet boundary, as expressed follow:

$$\begin{cases} g(\boldsymbol{x}) = \bar{\boldsymbol{u}}, \ l(\boldsymbol{x}) = 0, & \boldsymbol{x} \in \Gamma_D \\ g(\boldsymbol{x}) = 0, \ l(\boldsymbol{x}) > 0, & \boldsymbol{x} \notin \Gamma_D \end{cases}. \tag{15}$$

This procedure ensures that the imposed boundary conditions are consistently incorporated into the energy computation, thereby preserving the physical integrity of the solution. Although the loss function is evaluated on a single device, the PyTorch [40] framework provides native support for automatic gradient computation and distribution across all participating devices during the backpropagation process. This built-in functionality obviates the need for manual intervention, thereby streamlining the implementation of multi-GPU parallel training and enhancing both the scalability and robustness of the computational framework.

An interesting observation during the experimental process is that, due to the incorporation of spatial coordinates as input variables in the PINNs, accurate solutions can, under certain conditions, be achieved without the explicit enforcement of interface constraints when utilizing a single neural network. This phenomenon arises from the spatial proximity of nodes positioned along the interfaces of adjacent subdomains, whereby the same neural network tends to yield similar predictions for these closely situated nodes. This behavior is herein referred to as the weak spatial constraint (WSC) effect. However, the WSC effect is not universally beneficial. In scenarios where structural components on either side of a gap are expected to exhibit distinct deformation behaviors, the WSC effect may inadvertently neglect the discontinuity at the gap. Specifically, when the gap is sufficiently small, the network tends to enforce a continuous displacement field across the gap, thereby disregarding the gap's influence. In contrast, the enforced interface constraints-based DDM proposed in this study, which effectively mitigates this issue. By utilizing distinct neural network models for each subdomain, the WSC effect is inherently avoided. The specific details of this approach will be further explored in the following numerical examples.

## 3. Numerical experiments and discussion

This section evaluates the predictive performance of the proposed discrete physics-informed neural network (dPINN) framework with enforced interface constraints (EIC) under the domain decomposition method (DDM). The primary objective is to assess its capacity to decouple the weak spatial constraint (WSC) effect within a multi-GPU environment, in comparison to a conventional single-GPU setting, thereby evaluating both spatial independence and predictive fidelity. The parallel efficiency enabled by the domain decomposition strategy are also investigated. Unless otherwise specified, all numerical experiments are conducted on **a computing platform equipped with multiple NVIDIA RTX 4090 GPUs and an XX-core CPU**, with the material model defined by a Young's modulus of 3.0 GPa and a Poisson's ratio of 0.3.

## 3.1 Decoupled Weak Spatial Constraint

PINNs have emerged as a promising framework for solving PDEs, capable of approximating both globally continuous solutions or discrete pointwise values at prescribed spatial locations. However, due to the intrinsic dependence on spatial coordinates as input features, PINNs are inherently constrained by spatial alignment effects. Alleviating such limitations typically necessitates the design of more complex neural network architectures to capture highly nonlinear behaviors. In contrast, the DDM-dPINN with enforced interface constraints developed in this study inherently addresses these issues, offering a promising approach to improving the flexibility and accuracy of multi-subdomain solution strategies. Figure 4 illustrates the discretized models employed for numerical simulations, along with the corresponding boundary condition configurations. For enhanced visual clarity, the scales of the two discrete models are not normalized. Readers are advised to refer to the labeled dimensions for accurate interpretation. In Figure 4 (a), the overall structure represents a clamped domain with a geometric gap. The computational domain is partitioned into two subdomains, with the non-conforming mesh interface on the left side of the circular hole serving as the subdomain interface, while a physical gap is present on the right side. The second model in Figure 4 (b), represents a simplified hook structure, in which the non-conforming interfaces are positioned at the handle and the tip of the hook. Both displacement and external loading boundary conditions are imposed in a nodal form on the discretized model. Vertically downward forces are applied to all designated load nodes, resulting in a total resultant force of $3.6 \times 10^4$ kN in both discrete models.

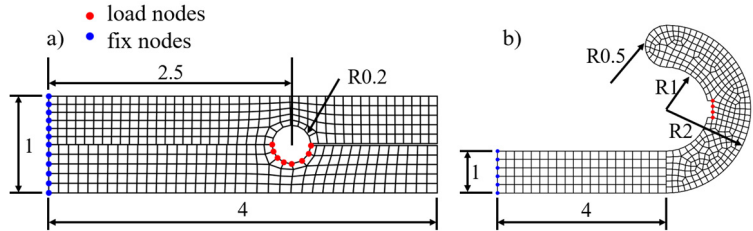

Figure 4 Illustration of a discrete model with boundary conditions considering a gap.

To investigate the presence of spatial weak coupling, two configurations of the dPINN framework are considered. The first configuration, referred to as the EIC-dPINN approach, employs a single neural network without DDM capabilities, where one network processes the inputs and outputs for all subdomains. The second configuration, referred to as the DDM-EIC-dPINN approach, utilizes two identical neural networks, with each network dedicated to its respective subdomain.

$$\begin{cases} u_{\theta s} = f^{EIC}(\mathcal{NN}(\mathcal{M}^{(1)} \cap \mathcal{M}^{(2)})) \\ u_{\theta m} = f^{EIC}(\mathcal{NN}^{(1)}(\mathcal{M}^{(1)}), \mathcal{NN}^{(2)}(\mathcal{M}^{(2)})) \end{cases}. \quad (16)$$

Results derived from non-conforming finite element simulations are employed as reference benchmarks for performance evaluation. To minimize the influence of confounding factors, both network architectures are designed with the same depth, and the width of the EIC-dPINN model is set to twice that of the DDM-EIC-dPINN to maintain a comparable number of trainable parameters. Enhanced interface constraints are imposed at the subdomain boundaries in both models. Each network comprises four hidden layers, with widths of 112 and 56 for the EIC-dPINN and DDM-EIC-dPINN, respectively. The training is performed using the Adam optimizer [41] with an initial learning rate of $1\times10^{-3}$, following a cosine annealing decay schedule without restarts over 20,000 epochs.

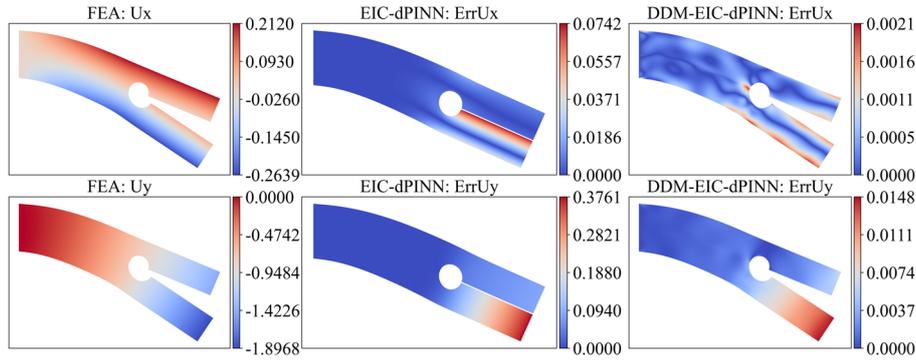

Figure 5 Visualization of weak spatial constraints: comparison of displacement fields.

The predicted displacement fields are presented in Figure 5, where all subplots display the actual deformed structures after loading. Notably, the color mappings represent different quantities across the columns. The first column illustrates the displacement field components computed via the non-conforming FEM, while the second and third columns depict the absolute displacement errors of the EIC-dPINN and DDM-EIC-dPINN models with respect to the FEM solution, respectively.

A notable observation is that the EIC-dPINN model fails to accurately capture discontinuities across spatial gaps. Owing to the use of a single neural network to approximate the solution across the entire domain, input coordinates that are spatially proximate, despite being physically separated, tend to exhibit similar output responses. As a result, the predicted displacement field by EIC-dPINN exhibits artificial continuity across the gap regions. In contrast, the DDM-EIC-dPINN model, which adopts a DDM architecture using two independent neural networks for each subdomain, effectively mitigates this issue. Each side of the spatial gap is handled by a distinct network, thereby inherently decoupling the prediction across the non-conforming interface and preserving the discontinuity. Regarding the distribution of errors, displacement errors in both the X and Y directions are predominantly localized in regions experiencing large deformations, particularly in the lower portion of the domain. The errors in the Y direction are relatively more significant. Nevertheless, the maximum absolute error reaches 0.0148, with the corresponding relative error remaining as low as 0.78%, thereby demonstrating that the prediction accuracy is effectively maintained.

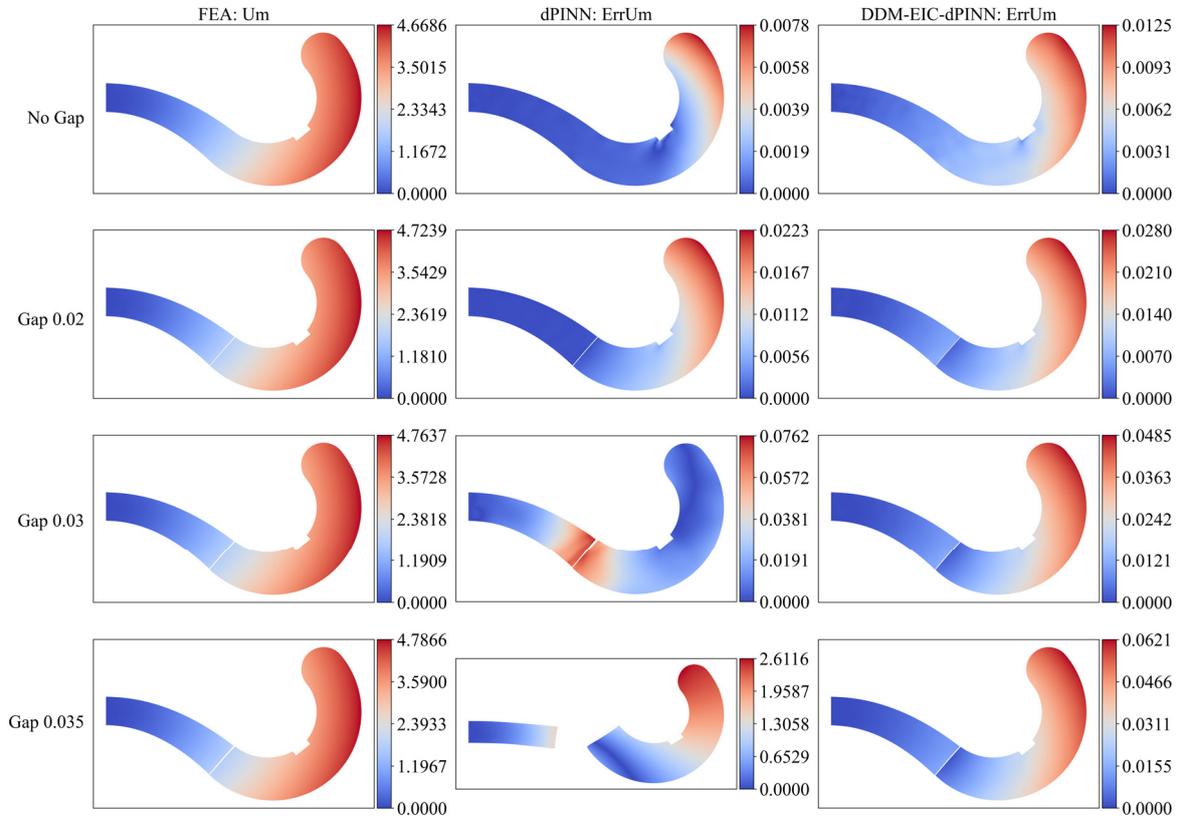

Figure 6 Displacement field-based illustration of enforced interface constraints.

It is worth noting that the WSC effect is not intrinsically detrimental. Experimental observations indicate that, under certain conditions, satisfactory predictive accuracy can be achieved without the explicit application of enhanced interface constraints by leveraging the WSC effect. To further investigate the positive contribution of WSC, the enhanced interface constraints were removed from the EIC-dPINN framework. As shown in Figure 6, the two subdomains are not fully connected, and additional examples with gaps width of up to 0.035 m were introduced. The figure illustrates the magnitude of the displacement field, and it can be observed that, in the absence of a gap, the dPINN achieves even higher predictive accuracy than the DDM-EIC-dPINN. This improvement may be attributed to the WSC effect partially offsetting the lack of explicitly enforced interface constrains, while the omission of displacement mapping operations facilitates faster convergence. As the gap gradually increases to 0.03 m, the predictive accuracy of dPINN progressively declines and becomes lower than that of the DDM-EIC-dPINN. At a gap width of 0.03 m, the maximum relative error reaches approximately 1.60%, with the predominant errors concentrated around the gap region. At this stage, the dPINN is nearing failure, as significant distortions in displacement are observed at the interfaces. As the gap increases further to 0.035 m, the impact of the WSC effect diminishes markedly, making it difficult to maintain displacement continuity across the subdomain interfaces based solely on the spatial proximity of input coordinates.

These findings suggest that, although the WSC effect can enhance prediction under favorable conditions, its effectiveness deteriorates markedly as the degree of discontinuity or geometric complexity increases. In contrast, the proposed DDM-EIC-dPINN framework effectively avoids the adverse effects of the WSC phenomenon and substantially improves the capability to address complex geometrical configurations. Furthermore, by applying reinforced interface continuity constraints, the framework ensures solution continuity across subdomain boundaries, maintaining high predictive accuracy even under challenging conditions. In addition to improving solution quality, the DDM approach enhances computational efficiency by enabling parallelization and localized training, the details of which will be discussed in the subsequent sections.

## 3.2 Improved Computational Efficiency via Domain Decomposition

### 3.2.1 Different problem scales

In this subsection, we further investigate the performance of the DDM, not only in terms of its inherent capability to decouple WSC effects, but also its potential for efficiency improvement through multi-GPU parallel training. As illustrated in Figure 4 a), there are three different levels of discretization of analytical models: a small-scale model with 152,156 elements, a middle-scale model with 809,679 elements, and a large-scale model with 3,821,324 elements. To evaluate the acceleration performance of DDM, multiple sub-networks were deployed on both single-GPU and multi-GPU configurations, and ten independent training runs were conducted for each scenario.

The experimental results, presented in Figure 7, demonstrate that multi-GPU parallel training consistently outperforms single-GPU training across all considered problem scales. For the small-scale problem, the average training time using a single GPU was 214.44 seconds, which was reduced to 195.28 seconds with the multi-GPU configuration, resulting in an approximately 10.35% reduction in training time. For the middle-scale problem, the training time on a single GPU was 899.97 seconds, while the multi-GPU setup decreased the training duration to 634.19 seconds, yielding a speedup of approximately 29.62%. In the case of the large-scale problem, the single-GPU training time of 4,285.88 seconds was reduced to 2,895.36 seconds when utilizing the multi-GPU approach, achieving a speedup of approximately 32.45%. This acceleration primarily stems from the ability of multi-GPU parallel training to distribute computational workloads, which would otherwise be executed sequentially on a single device, across multiple GPUs. This distribution significantly alleviates the computational burden on individual GPUs, leading to a reduction in overall training time. Particularly, as the problem size increases (middle and large scales), the computational demands grow exponentially, highlighting the limitations of single-GPU resources. Multi-GPU systems, on the other hand, leverage increased memory and computational resources, resulting in superior acceleration for large-scale tasks. Furthermore, as the problem size grows, the time cost associated with training models becomes increasingly dominant relative to the construction of the loss function, making the acceleration more pronounced. Additionally, as evidenced by the boxplots in the figure, while the overall acceleration effect is clear, the data variability in the multi-GPU training group is comparatively larger. This variability is likely attributable to several factors: during multi-GPU training, the model's predictions must be assembled across GPUs and utilized in the construction of the loss function, resulting in communication delays that induce fluctuations in the training time. Furthermore, minor workload imbalances among sub-networks across different GPUs may exacerbate the variability in training duration.

In conclusion, multi-GPU parallel training demonstrates superior acceleration performance across all problem scales, with the acceleration advantage becoming increasingly pronounced as the problem size grows. The observed variability may be mitigated through further optimization of communication strategies and improved load balancing. However, it should be noted that the primary focus of this work is on the proposed parallel training framework for DDM-EIC-dPINN, and optimizations related to data access and memory efficiency fall outside the scope of the current study.

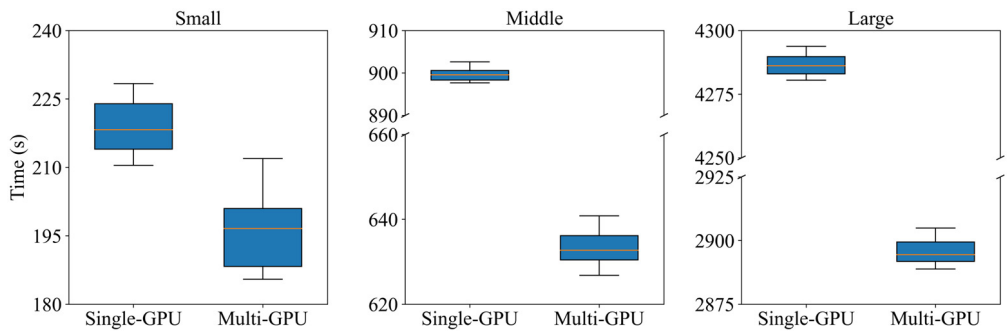

Figure 7 Training speed-up achieved through multi-GPU parallelization.

With the expansion of the problem scale from several hundred thousand to several million computational nodes, the associated increase in computational complexity is substantial. Nevertheless, by leveraging an efficient parallel training strategy, the model achieves a notable improvement in training efficiency while maintaining a high level of predictive accuracy. To rigorously assess the model's numerical performance across varying problem sizes, both the maximum relative error and the L2 norm error are adopted as quantitative measures of local and global prediction accuracy, respectively. The evaluation reveals that, as the computational model scales up, the maximum relative error in the X-direction displacement increases from 0.80% to 3.32%, with a corresponding rise in the L2 norm error from 0.50% to 1.51%. Similarly, the maximum relative error in the Y-direction increases from 0.78% to 2.44%, while the L2 norm error increases from 0.54% to 1.52%. Although a modest increase in error is observed with larger problem sizes, the overall accuracy remains within acceptable bounds, thereby preserving the reliability and validity of the numerical solution. Further comparative experiments reveal that increasing the number of training epochs for the Adam optimizer from 20,000 to 25,000 yields a notable improvement in predictive accuracy. Importantly, distributing the computational workload across multiple GPUs does not introduce any additional numerical error. As shown in Figure 8, the error curves obtained under single-GPU and multi-GPU configurations are completely identical, thereby confirming that the proposed parallelization strategy can substantially accelerate the training process without sacrificing solution accuracy. These findings affirm the robustness and scalability of the proposed framework, providing a reliable numerical foundation for large-scale physical simulations.

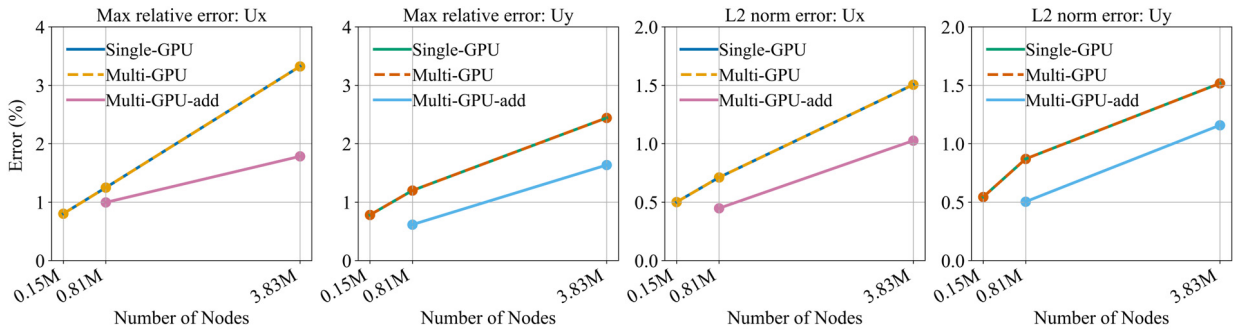

Figure 8 Evaluation of prediction errors in the displacement field, the suffix '–add' denotes an extended number of training epochs conducted with the Adam optimizer.

### 3.2.2 Multiple subdomains

As depicted in Figure 9, the original computational domain is divided into four non-overlapping subdomains to demonstrate the performance of the proposed DDM-EIC-dPINN framework in tackling problems with complex geometrical configurations. This domain decomposition strategy allows a geometrically complex global domain to be systematically transformed into a set of regular, structurally simpler subdomains. This reformulation not only streamlines mesh generation within each subdomain but also reduces the overall computational complexity. In addition, it significantly improves scalability by enabling efficient parallel computation across subdomains. In this experiment, the neural network architecture assigned to each subdomain is kept consistent with that used in the previous experiments.

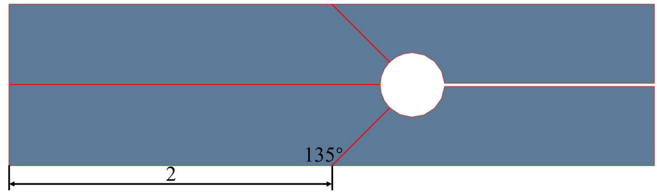

Figure 9 Geometry model of domain decomposition into multiple subdomains.

In the scenario involving decomposition into four subdomains, the overall mesh consists of approximately 0.24 million nodes. As depicted in Figure 10, the average training time on a single GPU is recorded as 361.35 seconds, while multi-GPU execution reduces this to 273.02 seconds, corresponding to a performance speedup of 24.44%. These results clearly indicate that the DDM framework continues to deliver acceleration benefits even in more complex multi-subdomain settings, with improvements becoming more pronounced as additional computational resources are leveraged. The observed speedup stems from the increased degree of parallelism, which offsets the sequential computational burden inherent in single-GPU training. However, this configuration introduces a moderate increase in total computational cost across both single- and multi-GPU settings. This overhead arises primarily from the larger number of neural networks that must be independently trained within each subdomain, as well as increased latency associated with inter-process data synchronization. Moreover, the number of enforced interface continuity constraints rises to three, further contributing to the computational workload. Despite the increase in total training cost, the multi-subdomain approach remains valuable in practice. In many scenarios, domain decomposition into multiple subregions is essential—not necessarily to minimize total computational time, but to simplify mesh generation for complex geometries and to enable more manageable subdomain structures. In such cases, the DDM framework effectively mitigates the associated computational overhead, maintaining its efficiency and scalability advantages.

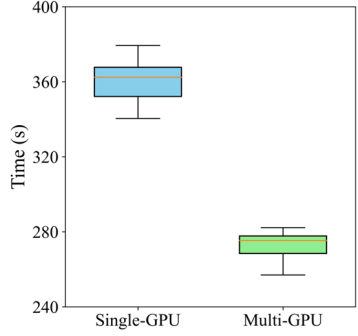

Figure 10 Performance acceleration of parallel training with multiple subdomains.

Increasing the number of interface constraints from one to three adds complexity to the training process of the DDM-EIC-dPINN framework. With the subnetwork architecture and training hyperparameters kept unchanged, As depicted in Figure 11, the maximum relative error in the X-direction rises to 3.42%, and to 2.91% in the Y-direction. As shown in earlier experiments, such errors can be reduced by increasing the number of training epochs, demonstrating that the accuracy of the method remains adjustable through further training. In addition, the predicted displacement error distributions obtained using both single-GPU and multi-GPU configurations remain nearly identical. This result confirms that the parallel domain decomposition strategy does not affect the prediction accuracy of the DDM-EIC-dPINN model. Instead, it improves computational efficiency and reduces hardware demands without introducing additional error.

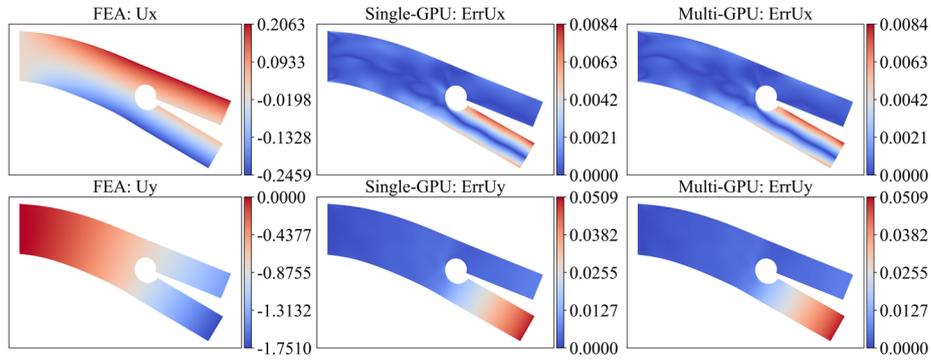

Figure 11 Evaluation of prediction error for DDM-EIC-dPINN with multiple subdomains.

The experiment demonstrates that the DDM-EIC-dPINN framework is capable of effectively handling domain decompositions involving two or more subdomains. However, an increase in the number of subdomains leads to a corresponding rise in both the total computational cost and the training complexity. Nevertheless, this approach proves advantageous for partitioning complex geometries into simpler subdomains. Hence, the number of subdomains should be minimized, provided that it does not significantly compromise the complexity of mesh generation.

## 3.3 Three-dimensional problem application

To further evaluate the reliability and applicability of the DDM-EIC-dPINN framework in complex 3D settings, a 3D computational experiment is conducted. As depicted in Figure 12, the geometric model considered in this study is a Hollow tubular structure constructed by sweeping a rectangular cross-section along a straight path until it transitions into a circular ring-shaped cross-section. This results in a geometry that combines both cross-sectional variation and spatial complexity. The rectangular end of the structure is fully constrained to impose fixed boundary conditions, while an external load is applied to the circular end. The resultant force, composed of components in the Y- and Z-directions, has a total magnitude of $6.4 \times 10^3$ kN for each component. As a result of the transition to a three-dimensional computational domain, the input and output dimensions of the neural network are extended from 2 to 3, whereas all other network configurations and training hyperparameters remain consistent.

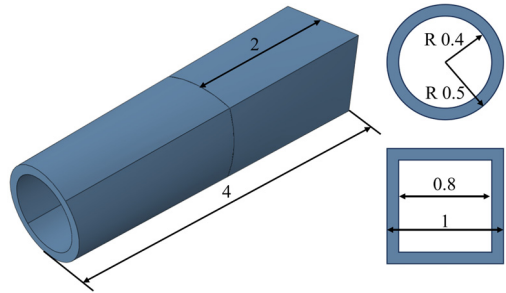

Figure 12 Geometry model of 3D example.

As illustrated in Figure 13, the DDM-EIC-dPINN framework maintains excellent predictive performance even in the three-dimensional setting. The top row illustrates the results obtained from FEM simulations, whereas the bottom row depicts the corresponding absolute error distributions of the DDM-EIC-dPINN predictions with respect to the FEM reference solution. The maximum relative errors in the displacement fields along the X-, Y-, and Z-axes are 0.94%, 0.86%, and 0.81%, respectively. The increased dimensionality does not impose substantial challenges on the training process, and the prediction accuracy remains consistently high. Moreover, the distribution of displacement errors shows no noticeable concentration near the subdomain interfaces, indicating that displacement continuity is effectively preserved across the entire domain. The smooth variation of error across the domain further suggests that the enforcement of interface continuity through hard constraints significantly enhances the continuity of the predicted physical fields.

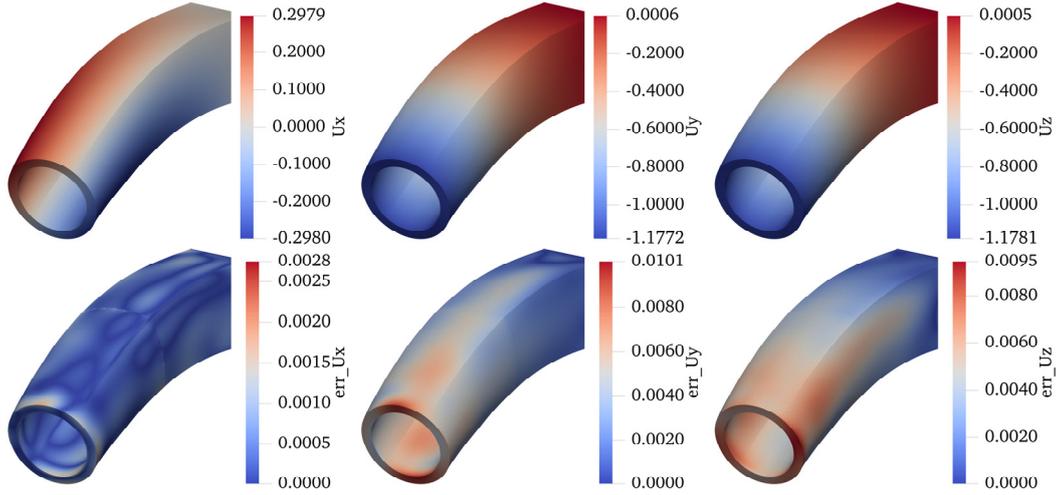

Figure 13 Evaluation of prediction error for DDM-EIC-dPINN under 3D scenario.

This advantage is particularly evident when compared to conventional methods in which continuity across subdomains is enforced indirectly through penalty terms added to the loss function, such as soft constraints on displacement and flux continuity. Notably, the additional training complexity introduced by explicitly imposed interface constraints remains significantly lower than that caused by the inclusion of multiple competing loss components. Such indirect formulations are often associated with unstable training dynamics, slower convergence rates, and diminished computational efficiency.

## 4. Conclusion

This study presents a discrete physics-informed neural network (dPINN) framework grounded in finite element discretization as the core modeling approach. In contrast to conventional collocation-based PINNs that approximate system energy via Monte Carlo integration, the proposed dPINN utilizes Gaussian quadrature for element-wise energy computation, thereby enabling a more accurate and physically consistent representation of the system energy. To enhance both the accuracy and scalability of the domain decomposition method (DDM), an enforced interface constraint (EIC) mechanism is introduced. This formulation enhances the accuracy of solutions across subdomain boundaries while alleviating the complexity typically associated with model training.

In two-dimensional configurations containing geometric discontinuities, such as gaps, it is observed that the inclusion of spatial coordinates as input features can induce unintended, weak spatial constrain (WSC) in the model's output. Although the WSC effect may exhibit both beneficial and detrimental characteristics, its uncontrolled nature undermines the robustness of the model. The proposed EIC formulation effectively decouples these spurious interactions by explicitly enforcing continuity conditions across interfaces. In contrast to conventional approaches that enforce interface continuity through soft constraints and penalty terms in the loss function, the proposed EIC formulation imposes continuity explicitly via hard constraints. This circumvents the challenges associated with tuning weight coefficients and avoids instability or bias in optimization arising from competing loss terms. As a result, the continuity of the predicted physical fields at subdomain boundaries is significantly improved without increasing the loss function's complexity.

Extensive numerical experiments on both two- and three-dimensional benchmark problems, validated against high-fidelity finite element simulations, confirm that the DDM-EIC-dPINN framework consistently delivers high predictive accuracy. Although a modest decline in accuracy is observed as the problem scale increases, this can be effectively mitigated by extending the number of training epochs. Furthermore, the proposed method offers a straightforward and flexible formulation for domain decomposition, enabling complex geometries to be partitioned into simpler, regular subdomains. The parallel training enabled by this strategy leads to substantial improvements in computational efficiency while preserving solution fidelity.

In summary, the DDM-EIC-dPINN framework constitutes a robust and scalable solution paradigm for complex geometrical domains. By enforcing physical field continuity through EICs and leveraging DDM-based parallel training, it substantially improves computational efficiency while alleviating hardware constraints.